\documentclass[pdflatex,sn-basic]{sn-preprint}

\jyear{2022}%

\theoremstyle{thmstyleone}%
\usepackage{bm}
\usepackage{subfigure}

\theoremstyle{thmstyletwo}%

\theoremstyle{thmstylethree}%

\newcommand{\pkg}[1]{{\fontseries{m}\fontseries{b}\selectfont #1}}

\begin{document}

\title[Mixture of Experts Distributional Regression]{Mixture of Experts Distributional Regression: Implementation Using Robust Estimation with Adaptive First-order Methods}


\author*[1,2,3]{\fnm{David} \sur{R\"ugamer}}\email{david.ruegamer@stat.uni-muenchen.de}

\author[1,3]{\fnm{Florian} \sur{Pfisterer}}\email{florian.pfisterer@stat.uni-muenchen.de}

\author[1,3]{\fnm{Bernd} \sur{Bischl}}\email{bernd.bischl@stat.uni-muenchen.de}

\author[4]{\fnm{Bettina}
\sur{Gr\"un}}\email{bettina.gruen@wu.ac.at}

\affil[1]{\orgdiv{Department of Statistics}, \orgname{LMU Munich},  \orgaddress{\city{Munich}, \country{Germany}}}

\affil[2]{\orgdiv{Department of Statistics}, \orgname{TU Dortmund}, \orgaddress{\city{Dortmund}, \country{Germany}}}

\affil[3]{\orgdiv{Munich Center for Machine Learning}, \orgaddress{\city{Munich}, \country{Germany}}}

\affil[4]{\orgdiv{Institute for Statistics and Mathematics}, \orgname{WU Vienna}, \orgaddress{\city{Vienna}, \country{Austria}}}


\abstract{In this work, we propose an efficient implementation of mixtures of experts distributional regression models which exploits robust estimation by using stochastic first-order optimization techniques with adaptive learning rate schedulers. We take advantage of the flexibility and scalability of neural network software and implement the proposed framework in \pkg{mixdistreg}, an \textsf{R} software package that allows for the definition of mixtures of many different families, estimation in high-dimensional and large sample size settings and robust optimization based on TensorFlow. Numerical experiments with simulated and real-world data applications show that optimization is as reliable as estimation via classical approaches in many different settings and that results may be obtained for complicated scenarios where classical approaches consistently fail.}




\maketitle

\section{Introduction}

Mixture models are a common choice to model the joint distribution of several sub-populations or sub-classes on the basis of data from the pooled population where the sub-class memberships are not observed \citep[for an introduction see][]{Mclachlan.2004}. Each sub-class is assumed to follow a parametric probability distribution such that the pooled observations are from a mixture of these distributions. Many applications of mixture models aim at estimating the distributions of the latent sub-classes and identifying sub-class memberships of the observations \citep{McLachlan.2019}. Mixture models have also been used in machine learning, e.g., for clustering \citep{Viroli.2019}, to build generative models for images \citep{Oord.2014} or as a hybrid approach for unsupervised outlier detection \citep{Zong.2018}. 

Different kinds of mixture models are used depending on the available data structure and the parametric model which is assumed for each sub-population or sub-class. In the following, we assume a supervised learning task and that regression models are used to model the sub-populations. This leads to mixture regression models which define a mixture of (conditional) models for the outcome of interest, where the mixture components are still unknown and thus considered latent variables. This model class is also referred to as mixtures of experts \citep{Gormley+Fruehwirth-Schnatter2019}. Mixtures of experts allow for the inclusion of covariates when modeling the mixture components as well as for the mixture weights. In the following, we consider a mixture of experts model where the regression models do not only include the mean parameter but also other distributional parameters that may depend on the covariates or features. In addition, we assume that the sub-population sizes vary with covariates (or features). This leads to the class of mixture of experts distributional regression models. 


\subsection{Mixture Models and Their Estimation} 

As for classical statistical regression models, the goal of \textit{mixtures of regressions} or \textit{mixture regression models} is to describe the conditional distribution of a response (or outcome), conditional on a set of covariates (or features). Mixtures of regressions have been first introduced by \citet{Quandt.1958} under the term \textit{switching regimes} where only two-component mixtures of linear regression models were considered, i.e., the number of sub-classes was fixed to two. This was extended to general mixtures of linear regression models by \citet{DeSarbo+Cron1988} who referred to this approach as a model-based version of \textit{clusterwise regression} \citep{Spaeth1979}. The extension to mixtures of generalized linear regression models was proposed in \citet{Wedel+DeSarbo1995}. Aiming for a setting beyond mean regression \citep{Kneib2013}, a distributional regression setting can be used such as generalized additive models for location, scale and shape (GAMLSS; \citealt{Ribgy+Stasinopoulos2005}) which also allow for non-linear smooth relationships between covariates and the distributional parameter of interest \citep{Stasinopoulos+Ribgy+DeBastiani2018}.  


Mixture models can be estimated using various techniques, with the expectation-maximization (EM) algorithm based on the maximum-likelihood principle being the most prominent one. Other approaches include Bayesian methods such as MCMC algorithms with data augmentation \citep{Diebolt+Robert1994}.
An alternative way of model specification and estimation was proposed by \citet{Bishop.1994} who introduced mixture density networks (MDNs). MDNs use a mixture of experts distributional regression model specification. But the relationship between the covariates (or features) and the mixture weights or the distributional regression models is learned using a neural network. The training of MDNs is done using highly optimized stochastic gradient descent (SGD) routines with adaptive learning rates and momentum. These procedures have proven to be very effective in the optimization of large neural networks with millions of parameters and complex model structures. While often advantageous in terms of prediction performance, MDNs lack interpretability as the relationship between inputs (features) and distribution parameters is modeled by a deep neural network.


\subsection{Our Contribution}

\paragraph{Novel Modeling Approach}
In this work, we combine the ideas of interpretable mixtures of regression models and MDNs to allow for a mixture of experts distributional regression models in a very general setup. In particular, this approach enables modeling mixtures of sub-populations where the distribution of every sub-population is modeled using a distributional regression. Predictors of every distribution parameter in every sub-population can be defined by linear effects or (tensor product) splines and thereby allow not only for complex relationships between the features and the distributions' mean but also other distribution characteristics such as scale or skewness. Furthermore, the sub-population sizes may vary with features in a flexible way using again a combination of linear effects as well as (tensor product) splines.

\paragraph{Robust Estimation}
Estimating mixture regression models within a maximum-likelihood framework on the basis of the EM algorithm works well for smaller problems. However, higher-dimensional settings where the number of parameters is similar to or exceeds the number of observations are often infeasible to estimate. In fact, the convergence and stability of classical algorithms are more and more adversely affected by an increase in model complexity. Maximum-likelihood estimation of distributional regression models itself induces a non-convex optimization problem in all but special cases. Thus extending mixtures of mean regression models to mixtures of distributional regression models further complicates the optimization. 

In this work, inspired by MDNs, we suggest and analyze the usage of stochastic first-order optimization techniques using adaptive learning rate schedulers for (regularized) maximum-likelihood estimation. Our results show that this approach is as reliable as estimation via classical approaches in many different settings and even provides estimation results in complex cases where classical approaches consistently fail. 

\paragraph{Flexible and Scalable Implementation}
Common implementations of EM optimization routines are limited in their flexibility to specify a mixture of (many) potentially different distributions but in general focus on the case where all components have a parametric distribution from the same distributional family. Another (computational) limitation of existing approaches is faced for large amounts of data (observations) as methods usually scale at least quadratic in the number of observations. On the other hand, SGD optimizers from the field of deep learning are trained on mini-batches of data allowing large dataset applications and the use in a generic fashion for all model classes. We take advantage of this flexibility and scalability and implement the fitting of the proposed model class using \pkg{mixdistreg}, an \textsf{R} \citep{R} software package based on the \textsf{R} package \pkg{deepregression} \citep{rugamer2021deepregression} that allows for the definition of mixtures with components from many different distributional families, estimation in high-dimensional and large sample size settings and robust optimization based on TensorFlow \citep{TensorFlow}.  

\paragraph{Summary of our Approach and Overview on the Paper Structure}
Our framework unites neural density networks \citep{Magdon.1998} with (distributional) regression approaches, extends MDNs by incorporating penalized smooth effects, and comprises various frameworks proposed in the statistical community such as \citet{Leisch.2004, Gruen.2007, Stasinopoulos.2007} to estimate mixtures of linear, generalized linear, generalized additive, or distributional regression models. In contrast to many existing approaches, our framework further allows to estimate the mixture weights themselves on the basis of an additive structured predictor. We refer to this combination of the model class of mixtures of experts distributional regression with structured additive predictors and the estimation using neural network software as the \emph{neural mixture of experts distributional regression} (NMDR) approach.

\noindent The remainder of this paper is structured as follows. In Section~\ref{sec:meth}, we present our model definition. In Section~\ref{sec:est}, we introduce the architecture for SGD-based optimization in neural networks and discuss penalized estimation approaches including mixture sparsification.
We then demonstrate the framework's properties using extensive numerical experiments in Section~\ref{sec:numExp} and its application to real-world data in Section~\ref{sec:application}. We conclude with a discussion in Section~\ref{sec:discuss}.

\section{Methodology} \label{sec:meth}

Our goal is to model the conditional distribution of ${Y} \mid \bm{x}$ where ${Y}$ is the univariate outcome (or response) of interest. We assume that $Y \mid \bm{x} \sim \mathcal{F}$, where $\mathcal{F}$ is a mixture of parametric distributions $\mathcal{F}_m, m \in \{ 1,\ldots,M \} =: \mathcal{M}$. These distributions, in turn, depend on unknown parameters $\bm{\theta}_m(\bm{x}) \in \mathbb{R}^{k_m}$ that are influenced by a set of features (or covariates) $\bm{x} \in \mathbb{R}^p$. Furthermore, the non-negative mixture weights are also allowed to depend on features (or covariates) $\bm{x} \in \mathbb{R}^p$ such that $\pi_m(\bm{x})$ for all $m=1,\ldots,M$ with $\sum_{m=1}^M \pi_m(\bm{x}) = 1$.

\subsection{Model Definition}

For an observation $y$ in (a suitable subset of) $\mathbb{R}$ constituting the support from the conditional distribution of $Y\mid \bm{x}$, we define the conditional density by
\begin{equation} \label{eq:model}
    f_{Y\mid\bm{x}}(y \mid \bm{\vartheta}(\bm{x})) = \sum_{m=1}^M \pi_m (\bm{x}) f_m(y\mid\bm{\theta}_m(\bm{x})),
\end{equation}
i.e., by a mixture of density functions $f_m$ of the distributions $\mathcal{F}_m$. Each component density $f_m$ has its own $k_m$ distribution parameters $\bm{\theta}_m = (\theta_{m,1}, \ldots, \theta_{m,k_m})^\top$. $\pi_m \in [0,1]$ are the mixture weights with $\sum_{m=1}^M \pi_m = 1$.  The vector $\bm{\vartheta}= (\bm{\theta}^\top, \bm{\pi}^\top)^\top \in \mathbb{R}^{K+M}$ with $K=\sum_{m=1}^M k_m$ comprises all distribution parameters $\bm{\theta} = (\bm{\theta}_1, \ldots, \bm{\theta}_M)^\top$ and all mixture weights $\bm{\pi} = (\pi_1,\ldots,\pi_M)^\top$. 

Each of the parameters $\vartheta_j$ in $\bm{\vartheta}$ is assumed to depend on the features $\bm{x}$ through an additive predictor $\eta_j, j=1,\ldots,K+M$. For the distribution parameters $\bm{\theta}$  a monotonic and differentiable function $h_j, j=1,\ldots,K$ is assumed to provide a map between the additive structured predictor and the distribution parameter, i.e.,
$\vartheta_j = h_j(\eta_j(\bm{x}))$. The parameter-free transformation function $h_j$ ensures the correct domain of each $\vartheta_j$. E.g., if $\vartheta_j$ represents a scale parameter the function $h_j$ ensures that $\vartheta_j$ is positive while the additive predictor $\eta_j$ may take arbitrary values in $\mathbb{R}$.

For the mixture weights $\bm{\pi}$ a single monotonic and differentiable function $h_{K+1}$ is assumed to map the $M$ additive predictors to the $(M-1)$-dimensional simplex, i.e., $h_{K+1}$ maps $\mathbb{R}^M \to [0,1]^M$ under the condition that the sum of the weights is 1. This links the last $M$ additive predictors $\bm{\eta}_{\pi} := (\eta_{K+1},\ldots,\eta_{K+M})^{\top}$ to the set of mixture weights $\bm{\pi}$. The most common choice in this respect for $h_{K+1}$ is the softmax function $$h_{K+1}(\bm{\eta}_{\pi}) = (\text{softmax}_1(\bm{\eta}_{\pi}), \ldots, \text{softmax}_M(\bm{\eta}_{\pi}))$$ with $$\text{softmax}_j(\bm{\eta}) = \frac{\exp(\eta_j)}{\sum_{l=1}^M \exp(\eta_l)}.$$ 
This implies
\begin{equation}
    \label{eq:pis}
\pi_m(\bm{x}_i) = \frac{\exp(\eta_{K+m}(\bm{x}_i))}{\sum_{l=1}^M \exp(\eta_{K+l}(\bm{x}_i))}, \quad \text{for } m \in \mathcal{M}.
\end{equation}
Effects in the additive predictors $\eta_{K+l}, l=1,\ldots,M$, in \eqref{eq:pis} are not identifiable without further constraints. We do not enforce any constraints during model training. Model interpretation is still possible in relative terms (e.g., using a log-odds interpretation). However, some constraints would need to be imposed if identifiable regression coefficients for the additive predictors are to be obtained.

For all parameters in $\bm{\vartheta}$, the additive structured predictors $\eta_j(\bm{x})$ ensure interpretability of the relationship between the parameter $\vartheta_j$ and the covariates. For example, if $\eta_{K+M}$ is a linear model, i.e., $\eta_{K+M}(\bm{x}) = \bm{x}^\top \bm{\beta}$, the regression coefficients $\bm{\beta}$ can be interpreted as linear contributions of each of the features to the logits of the mixture weight for the last mixture component $M$.

\subsection{Additive Predictor Structure}

The model \eqref{eq:model} relates all model parameters $\bm{\vartheta}$ to features $\bm{x}$ through additive predictors $\eta_j, j=1,\ldots,K+M$. As different densities $f_m$ and also different parameters $\bm{\theta}_m$ have potentially different influences on the conditional distribution of $Y \mid \bm{x}$, every parameter in $\bm{\theta}$ is defined by its own additive structured predictor $\eta_j$. Here we assume the additive predictors to have the following structure:
\begin{equation} \label{eq:pred}
    \eta_j = \beta_{0,j} + \bm{x}_{\mathcal{L}(j)}^\top \bm{\beta}_j + \sum_{l\in\mathcal{S}(j)} f_{l,j}(\bm{x}),
\end{equation}
where $\beta_{0,j}$ corresponds to the model intercept, $\bm{\beta}_j$ are the linear effects for pre-defined covariates $\bm{x}_{\mathcal{L}(j)}$ with $\mathcal{L}(j) \subseteq \{1,\ldots,p\} \cup \emptyset$ being a subset of all possible predictors, and $f_{l,j}$ are non-linear smooth functions for one or more covariates in $\bm{x}$ with $\mathcal{S}(j)$ being a (potentially empty) set of indices indicating the covariates with non-linear predictor effects. We assume that every $f_{l,j}(\bm{x})$ can be represented by (tensor product) basis functions $B_{l,j,o},o=1,\ldots,O$, taking one or several columns of $\bm{x}$ as input and mapping these onto the space spanned by the basis functions. Denoting $\bm{z}_{l,j} := \bm{B}_{l,j}(\bm{x}) = (B_{l,j,1}(\bm{x}),\ldots,B_{l,j,O}(\bm{x}))^\top \in \mathbb{R}^O$, the non-linear smooth terms can be written as $f_{l,j}(\bm{x}) = \bm{z}_{l,j}^\top \bm{\gamma}_{l,j}$ where $\bm{\gamma}_{l,j} \in \mathbb{R}^O$ are the corresponding basis coefficients for $\bm{z}_{l,j}$. 

In principle such a flexible specification may also be used for $\eta_{K+1}, \ldots, \eta_{K+M}$, i.e., for the additive structured predictors related to $\bm{\pi}$. However, these predictors are usually assumed to share one and the same additive structure, i.e., $\bm{x}_{\mathcal{L}(j)}$ and $\bm{z}_{l,j}$ with $l \in \mathcal{S}(j)$ are the same for all $j$ related to the $M$ mixture weights. We summarize all model coefficients of the linear terms by $\bm{\beta} = (\beta_{0,1}, \bm{\beta}_1^\top, \beta_{0,2}, \bm{\beta}_2^\top, \ldots, \beta_{0,K+M}, \bm{\beta}_{K+M}^\top)^\top$ and all model coefficients for representing the non-linear smooth functions by $\bm{\gamma} = (\gamma_{0,1}, \bm{\gamma}_1^\top, \gamma_{0,2}, \bm{\gamma}_2^\top, \ldots, \gamma_{0,K+M}, \bm{\gamma}_{K+M}^\top)^\top$. In the following, we summarize these two parameter vectors as $\bm{\psi} = (\bm{\beta}^\top, \bm{\gamma}^\top)^\top$.

\subsection{Model Log-Likelihood}

Based on model (\ref{eq:model}) and the structures imposed on the predictors in \eqref{eq:pred}, the negative log-likelihood of the model parameters $\bm{\psi}$ for $n$ independent observations $\bm{y} = (y_1, \ldots, y_n)^\top \in \mathbb{R}^n$  and their corresponding $n$ observed feature vectors $\bm{X} := (\bm{x}^\top_1, \ldots, \bm{x}^\top_n)^\top \in \mathbb{R}^{n \times p}$ is given by the sum of the negative log-likelihood contributions $\ell_i(\bm{\psi})$ of each observation $i=1,\ldots,n$:
\begin{equation}
\ell(\bm{\psi}) := \sum_{i=1}^n \ell_i(\bm{\psi}) = -\sum_{i=1}^n \log \left\{ \sum_{m=1}^M \pi_m(\bm{x}_i) f_m(y_i\mid\bm{\theta}_m(\bm{x}_i)) \right\}. \label{eq:emprisk}
\end{equation}
The negative log-likelihood can be rewritten as the negative sum of the exponentiated log-likelihoods of both the mixture weights $\pi_m$ and the component densities $f_m$ using the log-sum-exp (LSE) function: %
\begin{equation}
    -\sum_{i=1}^n \log \left\{ \sum_{m=1}^M \exp \left[ \log \pi_m(\bm{x}_i) + \log f_m(y_i\mid\bm{\theta}_m(\bm{x}_i)) \right] \right\}. \label{eq:emprisk2}
\end{equation}
In practice, formulation \eqref{eq:emprisk2} is often preferred over \eqref{eq:emprisk} as it is less affected by under- or overflow problems.

\subsection{Identifiability}

Identifiability is of concern for the mixture of experts distributional regression model because of the identifiability problems which could occur due to the mixture specification, due to the additive structured predictors, and due to the softmax mapping to the mixture weights.

\paragraph{Mixture Models}
For any mixture model, trivial identifiability problems may arise due to label switching and overfitting with either empty or duplicated components. In addition, also generic identifiability problems may occur for mixtures of distributions but also for mixtures of regressions. A detailed overview of identifiability problems in the finite mixture case is given in \citet{Fruehwirth.2006}. 

\paragraph{Additive Structured Predictors}

The components of the additive model with predictor structures of the form \eqref{eq:pred} are in general only identifiable up to a constant and also require further restrictions if both linear and non-linear smooth effects are defined for one and the same covariate. For example, $\eta_j = \beta_{0,j} + f_{1,j}(x_1)$ can be equally represented by $\tilde\beta_{0,j} + \tilde f_{1,j}(x_1)$ by defining $\tilde\beta_{0,j} = \beta_{0,j} + c$ and $\tilde{f}_{1,j}(x_1) = {f}_{1,j}(x_1) - c$ with $c\in\mathbb{R}$, i.e., by adding and subtracting a constant $c$ in both terms. We ensure the identifiability of non-linear smooth terms $f_{l,j}$ by using sum-to-zero constraints for all non-linear additive functions such as splines or tensor product splines. The identifiability of linear effects $x\beta$ in the presence of a univariate non-linear effect $f(x)$ must also be ensured. In this case, several different options exist \citep[see, e.g.,][]{Ruegamer.2020}. The most straightforward way is to use a non-linear effect $f(x)$ with a basis representation that includes the linear effect as null space (and hence, for enough penalization as discussed in Section~\ref{subsec:pen}, results in a linear effect). In this case, $\mathcal{L}(j)$ only consists of variables that are only modeled using a linear component (e.g., categorical effects) and $\mathcal{L}(j) \cap \mathcal{S}(j) = \emptyset$.

\section{Model Representation and Robust Estimation}\label{sec:est}

The observed data log-likelihood is, in general, not concave and thus difficult to optimize. We suggest to use optimizers from the field of deep learning by framing our model as a neural network. This enables the fitting of the full model class of mixture of experts distribution regression models with additive structure predictors in a straightforward way. Numerical experiments confirm that this choice -- when properly trained -- is not only more flexible and robust than EM-based optimization but also makes large dataset applications feasible due to mini-batch training. 

\subsection{Neural Network Representation}
\begin{figure*}
    \centering
    \includegraphics[page=6,  trim=1.5cm 2cm 2cm 0cm, width = 0.85\textwidth]{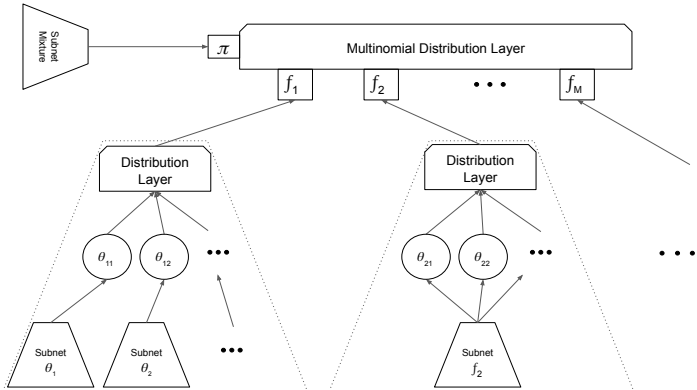}
    \caption{Example of an architecture. Smaller subnetworks (subnet) learn one or more parameters of a distribution which is defined in the respective distribution layer. For the first distribution in this example, each distribution parameter in $\bm{\theta}_1$ is learned through a separate network while the second distribution is learned by a network that outputs all parameters $\bm{\theta}_2$ together (e.g., used when $\bm{\theta}_2$ are constrained parameters such as probabilities). Each distribution layer thus corresponds to a distributional regression model. The mixture model is then defined by an additional subnetwork that learns the mixture weights $\bm{\pi}$ as well as by the $M$ learned distributions $f_1, \ldots, f_K$.}
    \label{fig:arch}
\end{figure*}

Models described in \eqref{eq:model} and \eqref{eq:pred} can be represented as neural networks in the following way. An exemplary architecture is depicted in Figure~\ref{fig:arch}. The network architecture implementing model (\ref{eq:model}) defines at most $K$ subnetworks, where each subnetwork models one or more additive structured predictors $\eta_{j}$ of a distribution parameter $\theta_{j}$. Using an appropriate parameter-free transformation function, these additive structured predictors are mapped to the distribution parameters and passed to a distribution layer \citep{TFP.2017}. Each distribution layer corresponds to a mixture component $f_m$ that is further passed to a multinomial or categorical distribution layer, modeling the mixture of all defined distributions. The mixture weights can either be directly estimated or also learned on the basis of input features using additive structured predictors in another subnetwork. A classical linear mixture regression combining $M$ linear regressions would, e.g., be given by $M$ subnetworks, each learning the expectation of a normal distribution and a mixture subnetwork that only takes a constant input (a bias) and learns the $M$ mixture weights. 

\subsection{Optimization Routines} \label{sec:comparison}

Representing the model \eqref{eq:model} and \eqref{eq:pred} as a neural network makes a plethora of optimization routines from deep learning readily available for model estimation. 
Various first-order stochastic optimization routines exist for neural networks. Most of these optimizers work with mini-batches $J_1,\ldots,J_B \subset \{1,\ldots,n\}$ of data, hence perform stochastic gradient descent (SGD). The update of parameters $\bm{\psi}^{[t]}$ in iteration $t\in\mathbb{N}$ and mini-batch $J_b$ is
\begin{equation}
    \bm{\psi}^{[t]} = \bm{\psi}^{[t-1]} - \nu^{[t]}  \frac{1}{\mid J_b\mid} \sum_{i\in J_b} \nabla_{\bm{\psi}} \ell_i(\bm{\psi}^{[t-1]}), 
\end{equation}
where $\bm{\psi}^{[0]}$ is some starting value, $\nabla_{\bm{\psi}} \ell_i(\bm{\psi}^{[t-1]})$ are the individual gradients of the negative log-likelihood contributions of the observations in the batch evaluated at the parameters of the previous iteration, and $\nu^{[t]}$ is a learning rate updated using a learning rate scheduler. We determine the starting value $\bm{\psi}^{[0]}$ with the Xavier uniform initialization scheme \citep{Glorot.2010}. For the update of $\nu^{[t]}$, we found the following learning rate schedulers useful for the optimization of mixture of experts distributional regression models: \textit{RMSprop} \citep{rmsprop}, \textit{Adadelta} \citep{adadelta}, \textit{Adam} \citep{adam}, and \textit{Ranger} \citep{ranger}. These optimizers in turn often come with hyperparameters such as their initial learning rate which need to be adjusted. We will investigate their influence in our numerical experiments section.

\subsection{Penalized Estimation} \label{subsec:pen}

Estimating the model class of mixtures of experts distributional regressions with additive structured predictors can benefit from a penalized log-likelihood specification. Such a penalization allows to control the smoothness or wiggliness of the non-linear smooth functions in the additive structured predictors and to induce sparsity in the estimation of the additive structured predictors as well as the mixture weights. 

\paragraph{Additive Structured Predictors}
In order to estimate suitable non-linear smooth additive structured predictors based on splines within the neural network, the respective coefficients in each layer may require regularization using appropriate penalties or penalty matrices. The smoothness or wiggliness of the non-linear additive structured predictor can be calibrated by either selecting a suitable dimension of the basis representation or by imposing a penalization on the respective coefficients in combination with a generous basis representation. Using the later approach, the penalized version of \eqref{eq:emprisk} is given by
\begin{equation}\label{eq:penrisk}
 \ell_{pen}(\bm{\psi}) =  \ell(\bm{\psi}) + \sum_{l \in \Lambda} \lambda_l \bm{\gamma}_l^\top \bm{P}_l \bm{\gamma}_l,
\end{equation}
with sets of index sets $\Lambda$ defining the weights that are penalized using a quadratic penalty with individual smoothing parameters $\lambda_l$ and individual quadratic penalty matrices $\bm{P}_l$ \citep[see, e.g.,][]{Wood.2017}. 
While estimating the tuning parameters is possible by including a more elaborated optimization routine in the neural network such as the Fellner-Schall method \citep{woodfasiolo2017}, we use the approach suggested by \citet{Ruegamer.2020} to tune the different smooth effects 
by relating $\lambda_l$ to their respective degrees-of-freedom $\text{df}_l$. This has the advantage of training the network in a simple backpropagation procedure with only little to no tuning, by, e.g., setting the $\text{df}_l$ equal to a pre-specified value for all $l \in \Lambda$. 



\paragraph{Entropy-based Penalization of Mixtures}

In order to penalize an excessive amount of non-zero mixture weights, we further introduce an entropy-based penalty for the mixture weights that can be simply added to the objective function in \eqref{eq:penrisk}:
\begin{equation}
    \ell_{ent}(\bm{\psi}) = \ell_{pen}(\bm{\psi}) - \xi \sum_m \pi_m \log \pi_m. \label{eq:entpen}
\end{equation}
The second part of \eqref{eq:entpen} is controlled by a tuning parameter $\xi \in \mathbb{R}^+_0$ and corresponds to the entropy induced by the (estimated) marginal mixture weights $\bm{\pi}$, i.e., in case the mixture weights depend on covariates $\bm{x}$ the mixture weights obtained when averaging over the observed $\bm{x}$. A large value of $\xi$ enforces the weight distribution to be sparse in the amount of non-zero elements in $\bm{\pi}$, while smaller values will result in an (almost) uniform distribution of $\bm{\pi}$. As the entropy is permutation invariant w.r.t.\ the ordering of the components in $\bm{\pi}$, this penalty is particularly suitable for mixtures that are only identified up to a permutation of the component labels. We investigate the effects of this tuning parameter $\xi$ in Section~\ref{sec:highdimmix}. While it is technically possible to allow feature-dependency of the conditional mixture weights which depend on $\bm{x}$ in \eqref{eq:entpen}, further research is required to investigate the effect of the entropy-based penalty in this case.


\section{Numerical Experiments} \label{sec:numExp}

We now investigate our framework in terms of predictive  and estimation performance. To this end, we first compare our neural mixture of experts distributional regression (NMDR) approach with EM-based optimization routines to demonstrate competitiveness with state-of-the-art procedures. For this comparison the model class considered is restricted to only contain constant mixture weights and linear predictors for the distributional parameters in order to facilitate the use of available implementations using EM-based optimization routines. Based on this simulation setup, we also perform an empirical investigation of different optimization routines in the neural network context to provide a practical guideline for users of our framework. We then evaluate our approach considering a mixture of generalized additive regression models with additional noise variables to demonstrate the framework's efficacy when including additive predictor structures. Finally, we investigate an overfitting mixture setting, in which we simulate situations where EM-based optimization fails and NMDR provides sparsity off-the-shelf. 

\paragraph{Evaluation Metrics}
We measure the estimation performance of both the regression coefficients and the mixture weights $\bm{\pi}$ using the root mean squared error (RMSE), the prediction performance using the predictive log-scores \citep[PLS;][]{Gelfand.1994} on an independent test dataset of the same size as the training dataset, and the differences between the true and estimated class memberships using the adjusted Rand index \citep[ARI;][]{hubert1985comparing} as well as the accuracy (ACC) by deriving the estimated class memberships based on the maximum a-posteriori probabilities for each of the mixture components. In order to deal with the problem of label switching for the non-label-invariant performance measures, we first determine the labeling which induces the accuracy-optimal assignment and then calculate these performance metrics using this labeling. The accuracy-optimal labeling is computed using the true class memberships (known in this case as data is simulated) and the estimated class memberships as induced by the mixture posterior probabilities estimated by each model. 

\subsection{Comparison with EM-based Optimization} \label{sec:comparisonEM}

\begin{figure}[ht]
    \centering
    \includegraphics[width = 0.8\columnwidth]{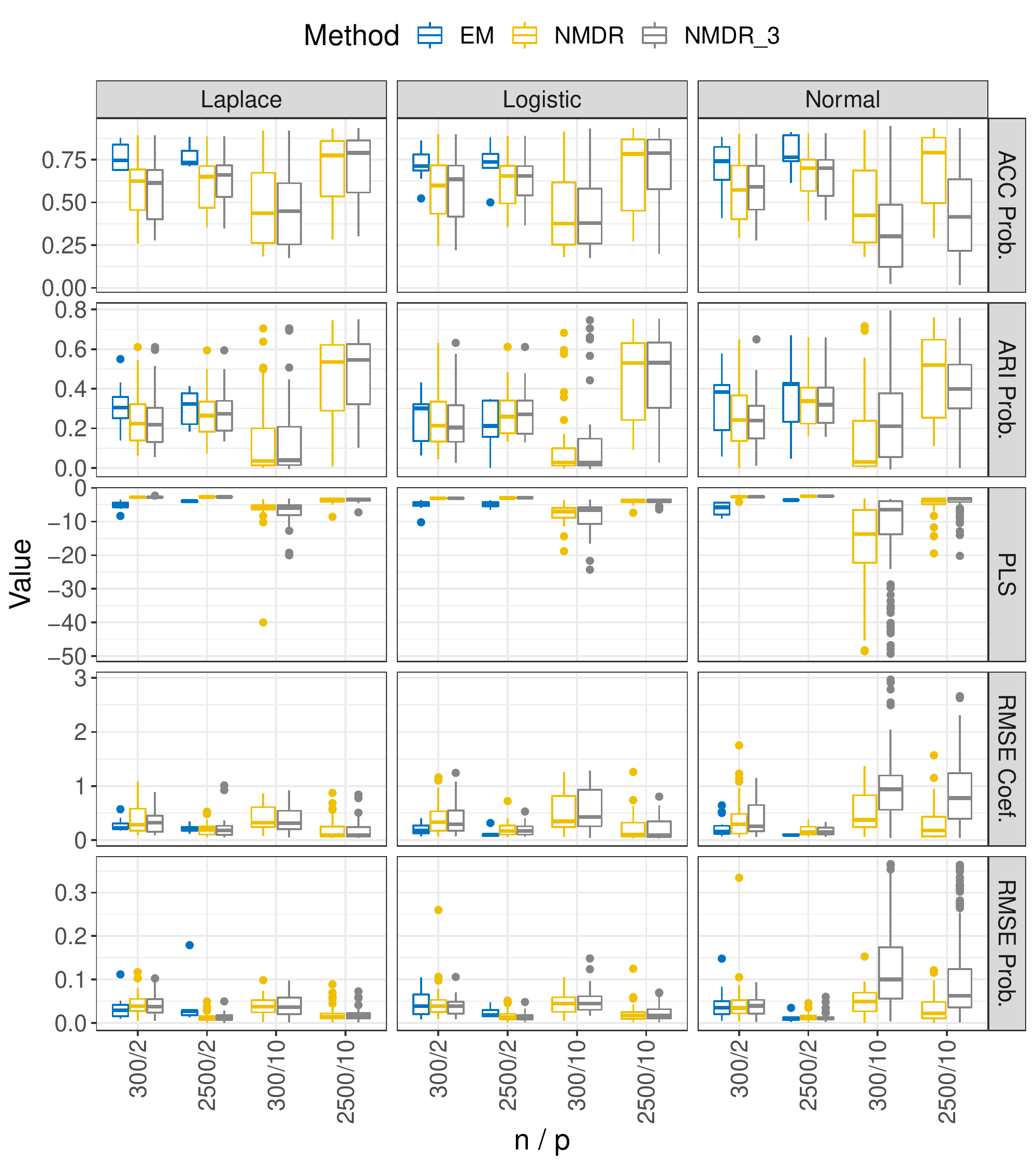}
    \caption{Comparison of EM-based optimization (EM), NMDR with one restart (NMDR) or with three restarts (NMDR\_3) for different distributions (columns), measures (rows), and combinations of $n$ and $p = p_m$. Boxplots contain all 10 runs and the four different settings for $M$ (i.e., in total 40 data points per boxplot). Missing boxplots for EM are due to missing solutions caused by missing values when comparing current results to a convergence threshold. RMSEs for coefficients $ > 3$ and PLS values $< -50$ are omitted to improve readability.}
    \label{fig:comparison_em}
\end{figure}

We first compare the NMDR framework to an EM-based algorithm implemented in the R package \texttt{gamlss.mx} \citep{gamlss.mx} allowing for mixtures of various distributional regressions. We use $n \in \{300, 2500\}$ observations, $M \in \{2,3,5,10\}$ identically distributed mixture components, either following a Gaussian, a Laplace, or a Logistic distribution, each defined by location and scale parameter, and mixture weights are randomly drawn such that the smallest weight is not less than 3\%. We use $p_m \in \{2, 10\}$ features for each distribution and distribution parameter in the mixture, and uniformly sampled regression coefficients from a $\mathcal{U}(-2,2)$-distribution. While we test \texttt{gamlss.mx} with a fixed budget of $20$ restarts, we compare these results to NMDR using $1$ and $3$ random initializations to assess the effect of multiple restarts. Each experimental configuration is replicated $10$ times. All fitted models in this simulation are correctly specified, i.e., they correspond to the data generating process.

\paragraph{Results} Results are visualized in Figure~\ref{fig:comparison_em} and yield four important findings. First, the EM-based approach only provides results in case $p_m = 2$. The EM-based approach is not able to converge to any meaningful solution in all settings with $p_m = 10$, whereas NMDR's performance is affected by the increased number of predictors, but still yields reasonable results for $n=2500$, also without restarts. Second, in the case $p_m = 2$, the EM-based approach provides in general a better classification performance than NMDR (as indicated by ACC Prob.\ and ARI Prob.). Third, the difference in estimation performance in case $p_m = 2$ between the EM-based and the NMDR approach is often negligible in terms of the RMSE between the estimated parameters. Fourth, while the induced regularization of the SGD-based routine induces a bias in the estimation and hence typically larger estimation errors, the predictive performance of NMDR is always better compared to the EM-based approach.

\subsection{Optimization Routines}

\begin{figure}[ht]
    \centering
    \includegraphics[width = 0.8\columnwidth]{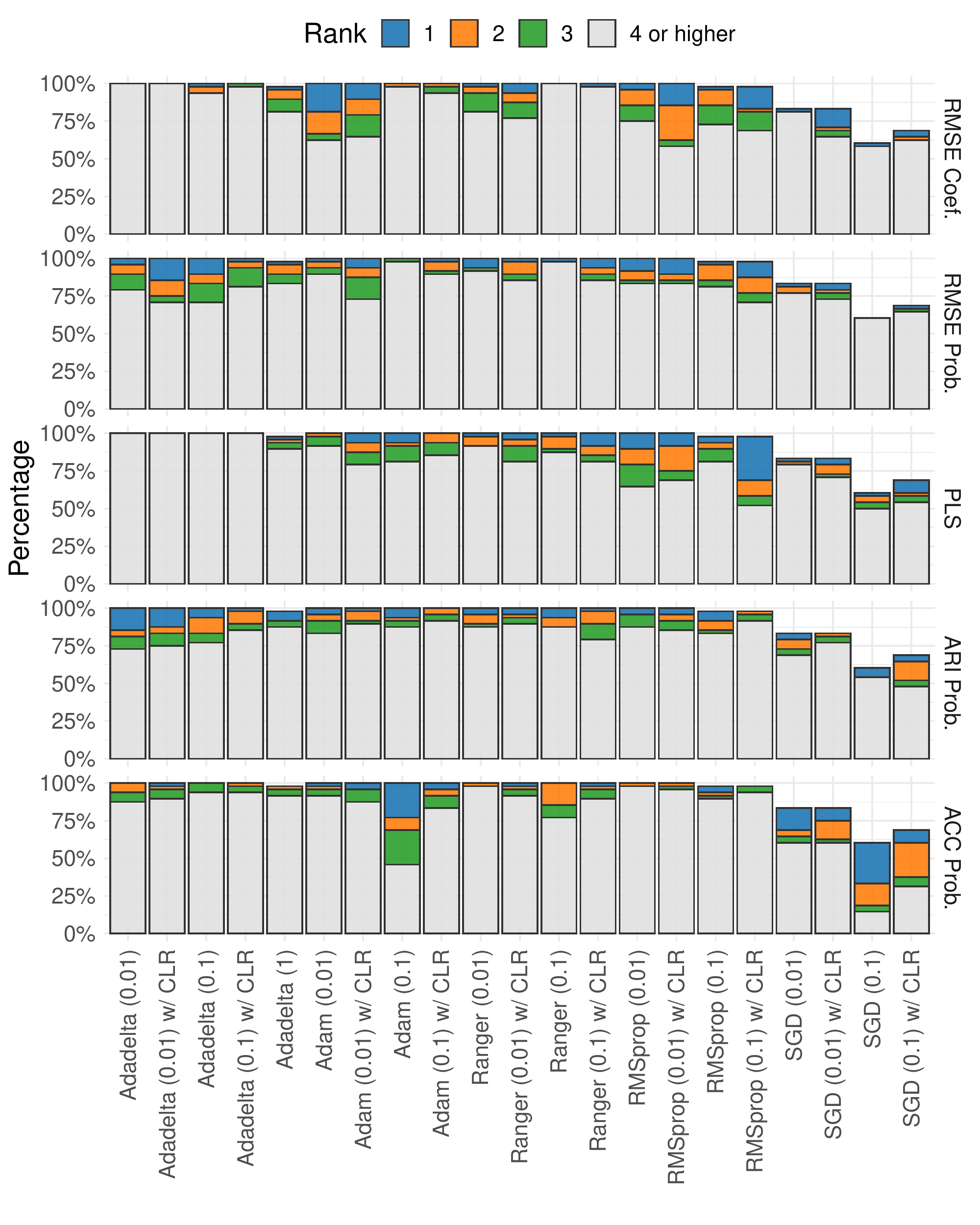}
    \caption{Comparison of various optimizers ($x$-axis; with learning rate in brackets) ranked by performance for different metrics (different rows), potentially with additional cyclic learning rate schedule (CLR), across simulation settings studied in Section~\ref{sec:comparisonEM}. Ranks are computed on performance averaged over 3 repetitions. Bars from SGD runs do not sum up to 100\% as some models diverged during optimization.}
    \label{fig:optimizers}
\end{figure}

In order to provide insights into the various optimizers' performance, we conduct a small benchmark study to assess the influence of the choice of an optimizer and to find a good default. Figure~\ref{fig:optimizers} visualizes the comparison based on the ranks for each optimizer across all 48 settings from Section~\ref{sec:comparisonEM}. 

\paragraph{Results} 
Figure~\ref{fig:optimizers} indicates that convergence problems are primarily encountered for SGD with a substantial amount of runs diverging during optimization. An overall performance assessment based on ranks indicates that the \textit{RMSprop} optimizer achieves the lowest overall rank. However, the figure highlights that in fact, no clear best optimizer emerges across all scenarios. The ranks obtained for the different optimizers also vary considerably with the performance criterion. 

Overall one might conclude, that both the RMSprop optimizer and the Adam optimizer perform in general well, also when used with their default settings. We note, however, that tuning the optimizer and its learning rate would in general also be beneficial in terms of both predictive performance and estimation quality. A further speed-up for some of these routines can be achieved by additionally incorporating momentum, which also proved to be effective in the optimization of additive models using boosting \citep{Schalk2022}.

\subsection{Mixture of Additive Regression Models} \label{sec:mixofdist}

\begin{figure}[ht]
    \centering
    \includegraphics[width = \columnwidth]{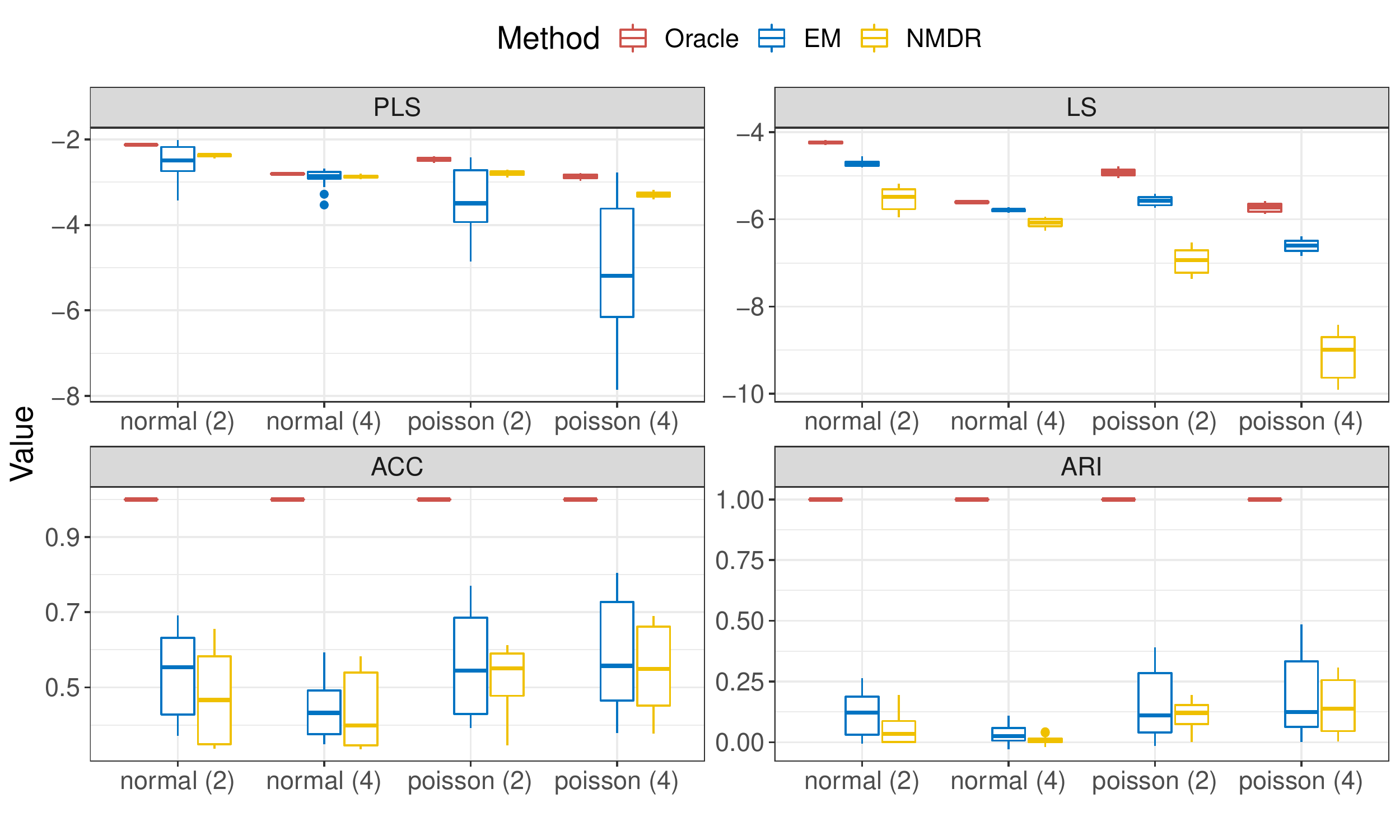}
    \caption{Comparison of average PLS and LS as well as ACC and ARI of a state-of-the-art implementation (EM), an oracle varying coefficient model with known class memberships, and our approach (NMDR) in different colors for the two distributions and two scales ($x$-axis).
    The boxplots contain the results for 10 replications over two settings for the mixture weights and two settings for the number of noise variables.}
    \label{fig:additive}
\end{figure}

Next, we investigate mixtures of mean regression models with non-linear smooth effects in the additive predictors of the mean distribution parameter. We generate $n=2500$ data points from $M = 3$ mixtures of Poisson or normal distributions with the additive structured predictor of the means defined by $\eta_1(\bm{x}) = \beta_0 + f_1(x_1) + f_2(x_2)$, $\eta_2(\bm{x}) = \beta_0 + f_2(x_1) + x_2$, and $\eta_3(\bm{x}) = \beta_0 + x_1 + f_3(x_2)$ with $\beta_0 = 0.5$, $f_1(x) = 2 \sin(3x)$, $f_2(x) = \exp(2x)$ and $f_3(x) = 0.2 x^{11} (10 (1 - x))^6 + 10 (10 x)^3  (1 - x)^{10})$. All covariates are independently drawn from a uniform distribution $\mathcal{U}(0,1)$. We model those effects using thin-plate regression splines from \citet{Wood.2017}. For Poisson data, we use $h(\cdot) = \exp(\cdot)$ and the identity for the Gaussian case. We vary $\bm{\pi}$ to be either $(1/3,1/3,1/3)$ 
or $(1/10, 3/10, 6/10)$ and 
add 3 or 10 noise variables 
that are also included in the model as non-linear smooth predictors for the expectation parameter. We use two different scale values $2$ or $4$, which either define the Gaussian variance in each mixture component or a multiplicative offset effect in the Poisson case. Our method is compared with a state-of-the-art implementation of mixtures of additive models using the \textsf{R} package \pkg{flexmix} \citep{Leisch.2004} and, as an oracle, a generalized additive model with varying coefficients for all smooth effects, where the class label (unknown to the other two approaches) is used as the varying parameter. For NMDR, the smoothing parameters are determined as described in Section~\ref{subsec:pen} via the respective degrees-of-freedom which are set equally for all smooths to $10$ for normal and $6$ for Poisson distribution.

\paragraph{Results} The comparison for all settings is depicted in Figure~\ref{fig:additive} showing the average log-score (LS; calculated on the training dataset using the estimated model parameters) and PLS, as well as the ARI and ACC.  The boxplots summarize all 10 simulation replications for the two different mixture weights and the two number of noise variables settings. Results suggest that our approach leads to better predictions measured by the PLS, but is inferior in terms of LS. The smaller LS values are possibly due to fewer data points to fit the model because of the need for a validation set, and due to the shrinkage induced by early stopping the procedure. The median performance of the clustering induced based on the estimated posterior probabilities is in general on par with the EM-based approach in terms of ARI and ACC with \pkg{flexmix}, while showing even slightly better performance in the Gaussian case.

\subsection{Misspecified Mixtures and Sparsity} \label{sec:highdimmix}

\begin{figure}[ht]
    \centering
    \includegraphics[width = 0.9\columnwidth]{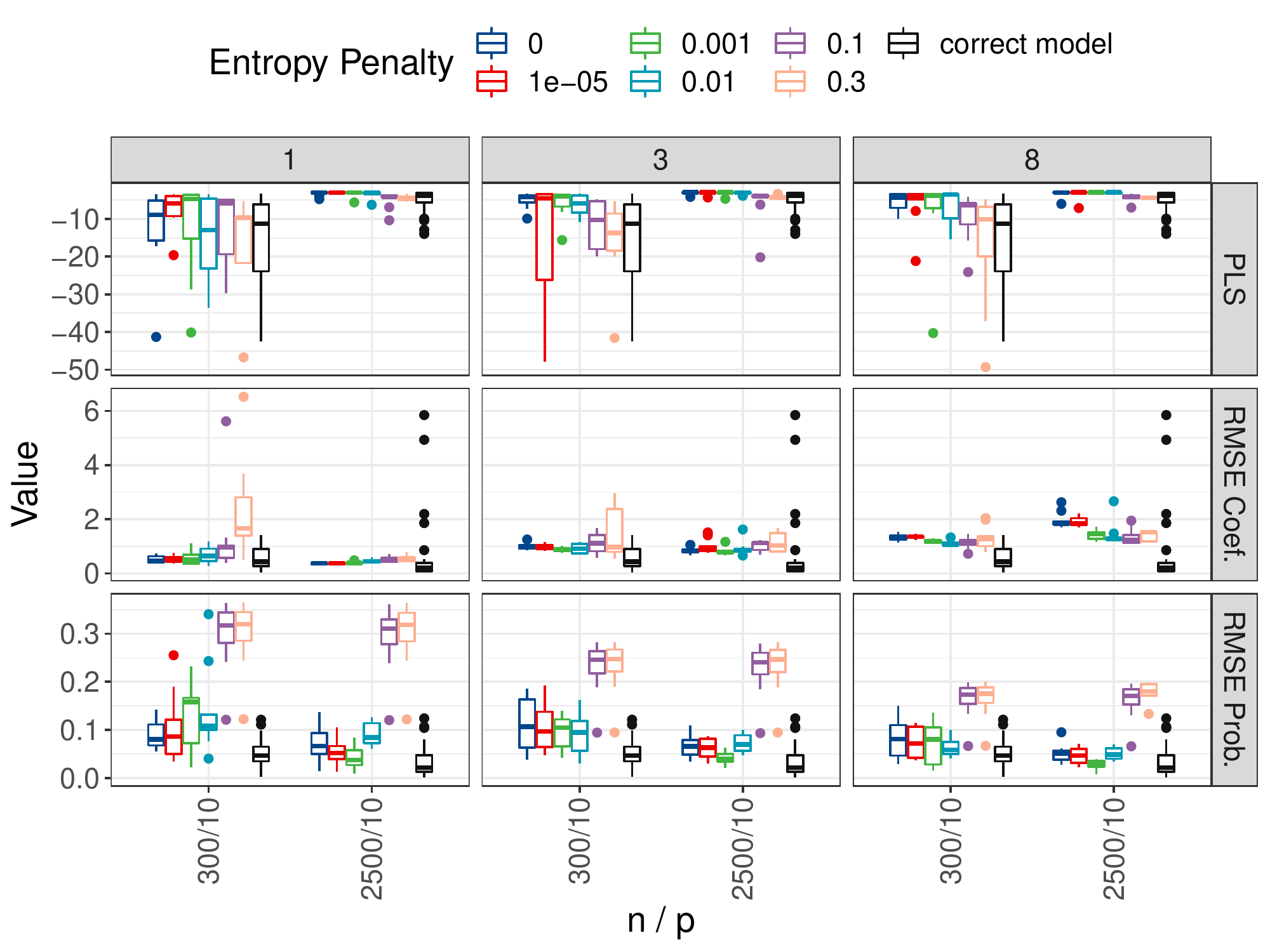}
    \caption{Model quality for misspecified models with specified mixtures $M^\dagger \in \{3, 5, 10\}$ (columns, counting the additional components) instead of actual $M=2$ mixtures and different goodness of fit measures (rows) for 10 replications (boxes). Colors correspond to different settings of $\xi$ or represent estimation results of the correct model (black).}
    \label{fig:add_mix}
\end{figure}

In this simulation, we use a normal mixture with $p_m = 10$ fixed linear predictors for each distribution and distribution parameter (mean and variance), where all features are again drawn from a standard normal distribution and regression coefficients from a uniform $\mathcal{U}(-2,2)$-distribution. The data is then generated using $M=2$ actual mixture components with $\pi_1$ drawn (independently of features) from a uniform distribution on the interval (0.06, 0.094) and $\pi_2 = 1 - \pi_1$ to ensure that the minimum value of both probabilities is at least 6\%. We then evaluate the estimation of mixture probabilities by NMDR for $n\in \{300, 2500\}$ when increasing the number of specified distributions $M^\dagger \in \{3, 5, 10\}$. To allow for sparsity in $\bm{\pi}$, we use the objective function $\ell_{ent}$ introduced in Section~\ref{subsec:pen}. For each scenario, 10 replications are performed. 

\paragraph{Results} Results for various settings of the entropy penalty parameter $\xi$ are depicted in Figure~\ref{fig:add_mix}. While setting $\xi$ to small values larger than zero can improve the predictive performance and even outperform the correctly specified model without additional distribution components, the bias induced by the penalty generally decreases the estimation performance. 

\begin{figure}[ht]
    \centering
    \includegraphics[width = 0.75\columnwidth]{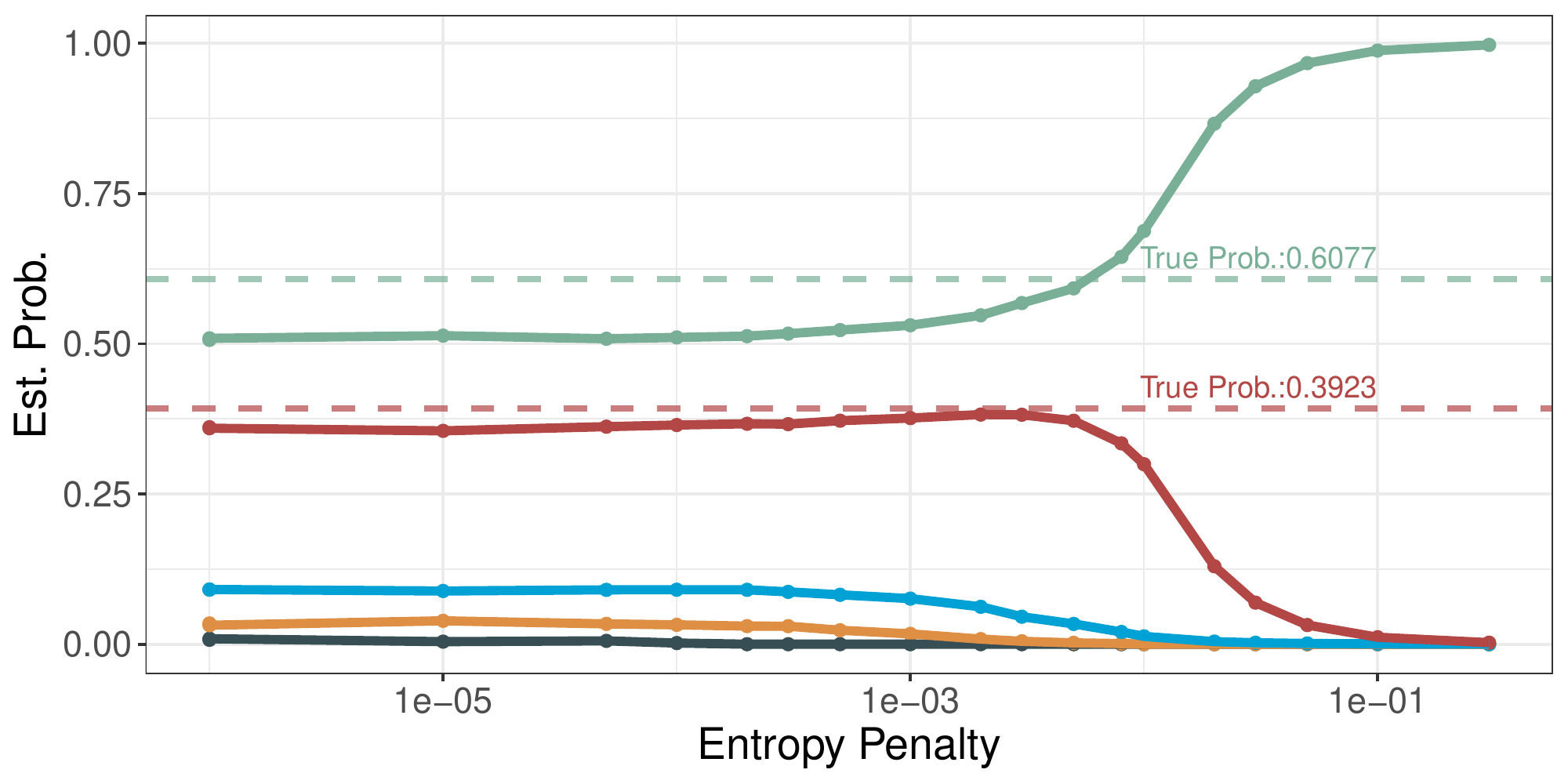}
    \caption{Coefficient path (estimated mixture probabilities) for different entropy penalties. A value of around 1e-02 yields the best trade-off between sparsity and estimation performance.}
    \label{fig:coef_path}
\end{figure}

We additionally investigate the coefficient path obtained from varying values of the penalty parameter $\xi$ for one simulated example. Results for the $n=2500, M^\dagger = 5$ setting are depicted in Figure~\ref{fig:coef_path}. $\xi$ is varied between 0 and 1 on a logarithmic scale. The true model has two non-zero probabilities $0.6077$ and $0.3923$ while the other $3$ entries in $\bm{\pi}$ are 0.



\section{Cell Cycle-regulated Genes of Yeast} \label{sec:application}

In order to demonstrate the flexibility of our approach, we investigate its application to the yeast cell cycle dataset from \citet{spellman1998comprehensive}. In this study, genome-wide mRNA levels were measured for 6178 yeast open reading frames (ORFs) for 119 min at 7 min intervals. We here analyze the subset of data where all 18 time points for the alpha factor arrest are available. The resulting longitudinal dataset consists of 80,802 observations of the standardized expression levels. A subset of this dataset was also analyzed using mixture models in \citet{Gruen.2011}. 

\paragraph{Distributional Mixture Regression}
As both the mean and standard deviation of the standardized expression levels of genes change over time, we apply a mixture of distributional regressions model where the mean $\mu$ and the standard deviation $\sigma$ of the normally distributed mixture components depend on time, i.e., $Y_t \sim \sum_{m=1}^M \pi_m \mathcal{N}(\mu_{m,t}, \sigma^2_{m,t})$ for $t\in [0,119]$. The additive structured predictors for these distribution parameters are defined as 
$$h^{-1}(\mu_{m,t}) = \beta_{0,m,1} + f_{m,1}(t) \quad \mbox{and} \quad h^{-1}(\sigma_{m,t}) = \beta_{0,m,2} + f_{m,2}(t),$$ 
where the non-linear smooth functions $f$ are modeled by thin-plate regression splines \citep{wood2003thin}. Previous approaches for modeling this dataset investigated the use of a mixture of mixed models, i.e., the inclusion of gene-specific random effects \citep{luan2003clustering, Gruen.2011}. We investigate here an alternative option for modeling this additional heterogeneity by allowing for time-varying standard deviations. We use $M=6$ which corresponds to the number of mixture components identified by \citet{spellman1998comprehensive}. 

\paragraph{Results}

In order to plot the estimated smooth effects together with the true observations, we first derive the component membership of every gene. As done in the E-step of mixture model approaches \citep[see, e.g.,][]{Gruen.2011}, we calculate the a posteriori probability for every gene to belong to component $m$ and then take the maximum of all components $1,\ldots,M=6$. For this application, observations were only assigned to 5 of the 6 assumed components. Note that due to the nature of our optimization routine, not all components necessarily contain at least one observation. 

\begin{figure}
    \centering
\includegraphics[width=\columnwidth]{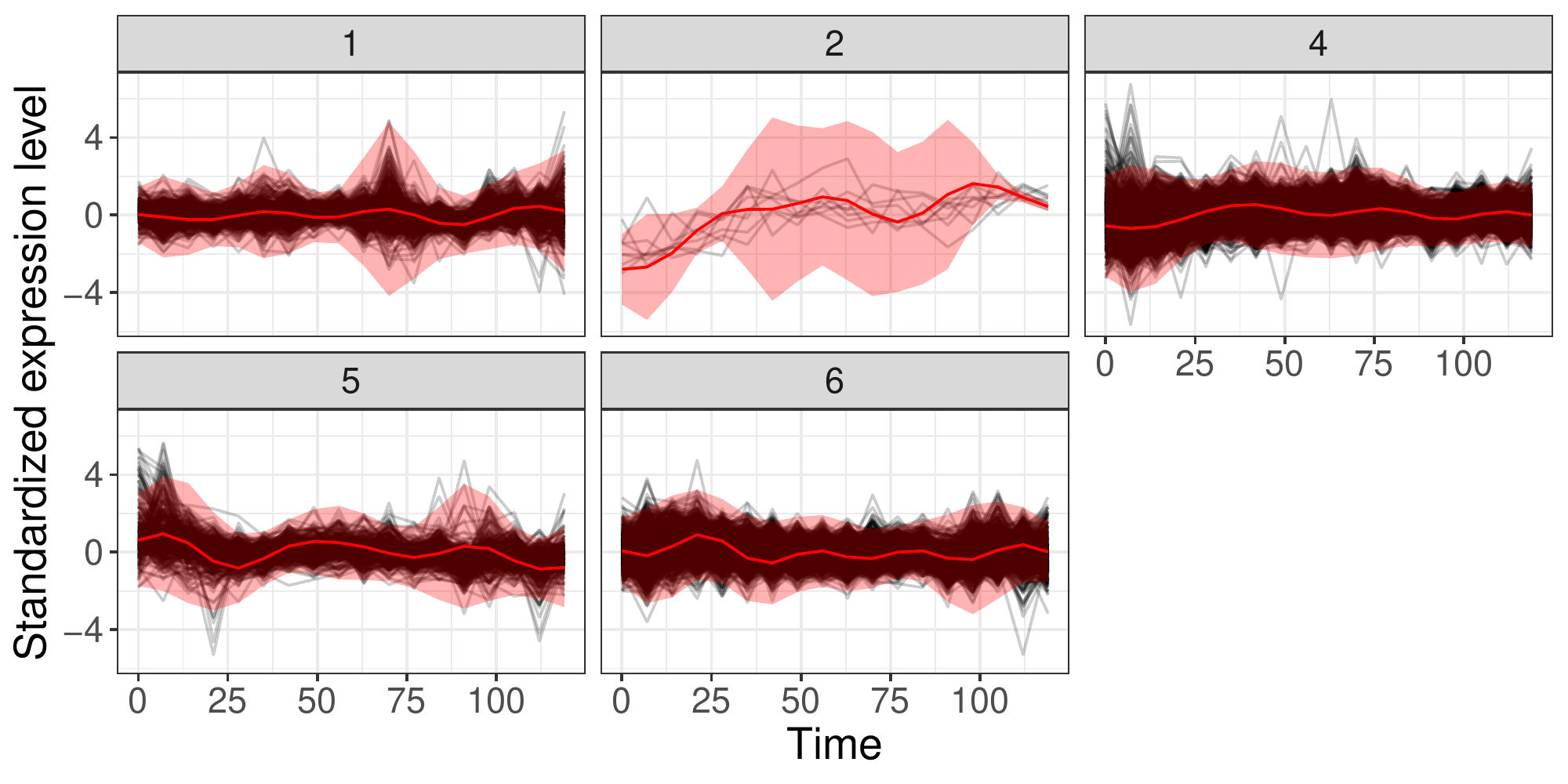}
    \caption{Trajectories of ORFs ($y$-axis) for all 4489 genes (black lines) over the course of the 18 time points ($x$-axis) with the estimated mean trend (red solid line) per component (facets) and uncertainty visualized by two times the estimated (time-varying) standard deviation (shaded red area).}
    \label{fig:yeast_1}
\end{figure}

Comparing the number of genes assigned to each cluster one sees that the most common component in our results is cluster 6 with 39,078 observations. Cluster 4 contains 25,722  observations, cluster 1 9,306, and cluster 5 6,552 observations. The least number of observations are assigned to cluster 2 which contains only 144 observations. 

Figure~\ref{fig:yeast_1} visualizes the results obtained in a panel-plot where in each panel the trajectories of the ORFs for all genes assigned to this cluster are shown together with the component-specific estimates of the time-varying means and standard deviations. The identified clusters clearly vary in showing either an initial decrease or increase in their means. In addition, one also sees that the standard deviations of the clusters vary over time with some clusters exhibiting a particularly large amount of heterogeneity at later time points.

\section{Outlook} \label{sec:discuss}

We have introduced the class of mixtures of experts distributional regression with additive structural predictors and investigated its embedding into neural networks for robust model estimation. Overall this leads to the neural mixture of experts distributional regression (NMDR) approach. We show that popular first-order adaptive update routines are well-suited for learning these mixture of experts (distributional) regression models and also highlight that the embedding into a neural network estimation framework allows for straightforward extensions of the general mixture model class and (regularized) maximum-likelihood estimation using optimization routines suitable also for big data applications due to mini-batch training. Using the proposed architecture for mixture of experts distributional regression, a possible extension of our approach is therefore the combination with other (deep) neural networks. This allows learning both the distribution components and the mixture weights either by (a) a structured model, such as a linear or additive model, (b) a custom (deep) neural network, or (c) a combination thereof. A similar approach has been investigated by \citet{fritz2022combining} using a zero-inflated Poisson model (i.e., a mixture including a point mass distribution) which includes both additive effects and a graph neural network in the additive predictor.

\backmatter







\section*{Declarations}

The authors declare that they have no known competing financial interests or personal relationships that could have appeared to influence the work reported in this paper.

\bibliography{sn-bibliography}


\end{document}